\documentclass[11pt]{article}
\usepackage[numbers,sort&compress]{natbib}

\usepackage[a4paper,margin=1in]{geometry}
\usepackage{amsmath,amssymb,amsfonts}
\usepackage{booktabs}
\usepackage{tabularx}
\usepackage{graphicx}
\usepackage{hyperref}
\usepackage{xcolor}
\usepackage{natbib}
\usepackage{enumitem}
\usepackage{caption}
\usepackage{array}
\usepackage{placeins}
\usepackage{float}
\usepackage{listings}
\usepackage{subcaption}
\usepackage{authblk}
	\usepackage{enumitem}

\usepackage{makecell}
\usepackage{siunitx}
\usepackage{threeparttable}

\usepackage{algorithm}
\usepackage{algpseudocode}

\hypersetup{
  colorlinks=true,
  linkcolor=blue,
  citecolor=blue,
  urlcolor=blue
}

\newcolumntype{Y}{>{\raggedright\arraybackslash}X}

\newtheorem{definition}{Definition}

\title{Machine-Coached Policy Revision in Adaptive Agent-Based  Regulatory Simulation:\\ A Controller-Level Contestability Layer}

\author{Roberto Garrone\\
Open University of Cyprus\\
\texttt{roberto.garrone@st.ouc.ac.cy}}

\date{}

\begin{document}

\maketitle

\begin{abstract}
	Policy-oriented agent-based models are increasingly used to study regulatory interventions in complex adaptive socio-technical systems. Recent adaptive ABM frameworks distinguish between static and adaptive agents, fixed and adaptive policies, and alternative controller designs. However, most diagnostic workflows remain ex post: trajectories are analysed after simulation, but the resulting evidence is not systematically fed back into the policy controller. This paper proposes a lightweight machine-coached policy-revision layer for adaptive agent-based regulation. The layer represents policy decisions as defeasible rules with explicit conflicts and priorities, generates explanations for controller actions, and allows diagnostic failures to be translated into rule additions, removals, or priority changes. The contribution is not a new optimal controller and does not claim formal guarantees for unrestricted machine coaching. Instead, it provides a simulation-compatible operationalization of controller-level contestability: policy decisions can be explained, challenged, revised, and re-evaluated in held-out simulation runs. A stylized emissions-regulation ABM is used as the experimental component. A controlled simulation experiment focuses on an over-conservatism failure in the VPVA regime. The predefined coaching template adds a relaxation rule to the symbolic controller, reducing over-conservatism recurrence under held-out seeds while preserving violation, overshoot, and volatility guardrails. The paper argues that machine coaching is best understood as a controller-level extension of explainable adaptive ABM, complementary to causal, information-theoretic, and trajectory-based diagnostics.
\end{abstract}


% ------------------------------------------------------------
\section{Introduction}
% ------------------------------------------------------------

Agent-based models (ABMs) are widely used to analyse policy interventions in complex socio-technical systems \cite{epstein1999,epstein2012,gilbert2008,conte2014,tesfatsion2006}. Their methodological value lies in representing heterogeneous agents, bounded rationality, local interaction, feedback, and emergent macro-level outcomes \cite{arthur1994,epstein1999,tesfatsionjudd2006}. This makes ABMs especially relevant in domains where aggregate policy effects cannot be reduced to a representative actor or a static response function \cite{epstein2012,gilbert2008}.

However, many policy-oriented ABMs still treat regulation as a fixed scenario parameter. A policy value is selected, the model is executed, and outcomes are compared across scenarios. This design is useful for counterfactual exploration, but it does not fully represent the adaptive structure of real regulatory systems \cite{garrone2025framework,garrone2026policy}. Regulators revise instruments in response to observed outcomes, firms or households adapt to incentives, and the resulting feedback changes the system being regulated \cite{tesfatsion2006,busoniu2008,zhang2021}.

Recent adaptive ABM work addresses part of this limitation by distinguishing four regimes: constant policy with constant agents (CPCA), constant policy with adaptive agents (CPVA), adaptive policy with constant agents (VPCA), and adaptive policy with adaptive agents (VPVA) \cite{garrone2025framework,garrone2026policy}. This taxonomy makes it possible to separate the effects of agent adaptation, policy adaptation, and their interaction. It also supports regime diagnostics based on scalar indicators, symbolic transition patterns, information-theoretic measures, and emergent trajectory clusters \cite{crutchfield1994,shalizi2001,garrone2026policy}.

Yet a gap remains. Diagnostics are usually computed after simulation. They help analysts interpret whether a controller violates a cap, behaves conservatively, oscillates near a boundary, or generates excessive volatility \cite{garrone2026policy}. But the diagnostic evidence is not normally incorporated into the policy controller itself. In other words, the model is diagnosable, but not yet revisable in a transparent and contestable way \cite{garrone2025framework,arrieta2020explainable}.

This paper proposes a controller-level machine-coaching layer for adaptive ABM. The key idea is simple: Diagnostic evidence should not only describe policy-controller failure; it should also provide structured grounds for revising the policy logic that produced the failure.

The proposed layer represents policy decisions through explicit symbolic rules, conflict relations, and priorities \cite{mccarthy1959,dung1995acceptability,nute1994defeasible}. When the controller selects an action, it also produces an explanation: which facts were observed, which rules fired, which rules were blocked, and which priority relation resolved the conflict \cite{arrieta2020explainable}. A coaching mechanism then translates diagnosed failures into rule-level revisions. These revisions may add a new exception, remove an invalid rule, or change a priority relation \cite{grosan2011rule,nute1994defeasible}.

Machine coaching is not presented as a substitute for causal analysis, sensitivity analysis,
information-theoretic diagnostics, or spatial decomposition \cite{crutchfield1994,shalizi2001,arrieta2020explainable}. Instead, it is positioned as a
specific controller-level layer for explaining and revising policy logic. The contribution is
threefold.

\begin{enumerate}[label=(\roman*)]
	\item It defines a lightweight rule-based policy controller for adaptive ABM regulation, with
	explicit rules, conflicts, priorities, and explanation logs.
	\item It introduces a template-constrained machine-coaching protocol that converts predefined diagnostic failures into auditable rule-base revisions.
	\item It implements and evaluates the protocol in a controlled emissions-regulation ABM,
	showing how a diagnosed over-conservatism failure can be translated into a relaxation rule
	and re-evaluated on held-out simulation seeds.
\end{enumerate}

This paper is methodological and derived from a broader PhD research programme on adaptive, explainable, and contestable ABM-based policy design \cite{garrone2025framework}.

% ------------------------------------------------------------
\section{Background and Motivation}
% ------------------------------------------------------------

\subsection{Adaptive ABM, policy regimes, and ex post diagnostics}

Let $s_t \in S$ denote the system state at time $t$, $X_t$ the vector of agent actions, $P_t$ a policy or control parameter, and $\zeta_t$ an exogenous disturbance. A generic ABM transition can be written as
\begin{equation}
	s_{t+1} = F(s_t, X_t, P_t, \zeta_t).
\end{equation}

Agents may be static or adaptive. Policy may be fixed or adaptive. Combining these two distinctions gives four dynamic regimes \cite{garrone2025framework,garrone2026policy}:
\begin{itemize}
	\item \textbf{CPCA}: constant policy, constant agents;
	\item \textbf{CPVA}: constant policy, variable/adaptive agents;
	\item \textbf{VPCA}: variable/adaptive policy, constant agents;
	\item \textbf{VPVA}: variable/adaptive policy, variable/adaptive agents.
\end{itemize}

This decomposition is useful because ``adaptiveness'' is not a single treatment. Agent adaptation alone may reduce emissions without ensuring compliance. Policy adaptation may track a constraint while producing frequent violations. Adaptive policy and adaptive agents may interact in ways that generate volatility, conservatism, or oscillatory boundary crossing \cite{garrone2026policy}.

Over a finite horizon, policy performance may be summarized by a bounded functional
\begin{equation}
	J(P;L) =
	\frac{1}{K}
	\sum_{t=T-K+1}^{T}
	\Phi(s_t),
\end{equation}
where $L$ denotes the agent learning or adaptation rule, $\Phi$ is a scalar evaluation function, and $K$ is the evaluation window. This formulation does not require convergence to equilibrium and remains meaningful under cyclic, drifting, or volatile dynamics \cite{busoniu2008,zhang2021,garrone2026policy}.

Adaptive ABM studies often compute scalar and structural diagnostics after simulation. Examples include: mean outcome; violation rate; overshoot; tracking error; policy volatility; transition entropy; motif diversity; conservatism gap; symbolic boundary-crossing patterns \cite{crutchfield1994,shalizi2001,garrone2026policy}.

These diagnostics can reveal that two controllers with similar mean performance behave very differently around a policy constraint. For example, one controller may track a regulatory cap closely but cross it frequently; another may avoid violations but remain far below the cap; a third may oscillate between excessive pressure and relaxation \cite{garrone2026policy}.

However, ex post diagnosis alone does not revise the controller. The analyst may observe that a controller is over-conservative, but the model does not explain which part of the policy logic generated that behavior, nor does it provide a reproducible way to change that logic. This creates a gap between interpretability and contestability \cite{arrieta2020explainable,garrone2025framework}.

\subsection{Related work}

The term machine coaching is used here in a restricted and operational sense, following Loizos' formulation of machine coaching as an interaction paradigm in which humans and machines externalize their reasoning in mutually understandable terms \cite{michael2019machinecoaching}. The proposed layer is not intended to cover every form of human--machine interaction, nor does it claim the full generality of open-ended natural-language coaching. Instead, it denotes a controller-level revision protocol in which diagnostic failures are linked to the explicit reasons that produced a policy action, and those reasons are revised through structured operations on the controller's symbolic rule base \cite{mccarthy1959,grosan2011rule,nute1994defeasible}. This restricted use is also consistent with later work on proxy coaches, where coaching is operationalized as an iterative exchange of explanations aimed at improving both machine conclusions and their acceptability to the coach \cite{markos2022proxycoaches}. In the present paper, the proxy coach is specialized as a predefined advice-template function that maps diagnostic failures to rule-base revisions.

First, the adopted approach differs from interactive machine learning and human-in-the-loop machine learning. In interactive machine learning, users typically provide labels, corrections, rankings, feature feedback, or demonstrations that help a model improve its predictive or decision performance \cite{amershi2014power}. In the present framework, the feedback object is not a label or a preferred trajectory. The feedback object is a defect in the controller's policy reasoning: for example, a rule was too broad, an exception was missing, or a priority relation produced an unacceptable regulatory response. The coach therefore acts on the explanation structure of the controller rather than directly on the simulated outcome.

Second, it differs from reinforcement learning from human feedback or human preferences. Preference-based reinforcement learning uses human judgments, often over trajectory segments, to infer or shape a reward model \cite{christiano2017deep}. The present controller does not learn a latent reward function from preference comparisons. It revises an explicit rule base whose conclusions, conflicts, and priorities remain inspectable before and after revision. The objective is not to produce an optimal policy by reward learning, but to make a policy controller contestable and revisable.

Third, the approach differs from post-hoc explainable AI. Post-hoc XAI methods usually explain a model after it has produced a prediction or decision, often without changing the model itself \cite{arrieta2020explainable}. In contrast, the proposed layer treats explanation as part of an adaptive revision loop. The explanation is not only an interpretive artefact; it identifies the rule, conflict, or priority relation that may be challenged and modified.

Fourth, the approach differs from algorithmic recourse. Recourse methods ask what changes would allow an affected subject to obtain a more favourable decision, often through counterfactual or intervention-oriented recommendations \cite{wachter2017counterfactual,karimi2020algorithmic}. The present framework does not primarily advise an agent how to change its state to obtain a different outcome. It revises the policy controller that generates regulatory actions. The object of intervention is therefore the decision logic of the regulator, not the feature vector of the regulated agent.

Fifth, the approach is related to, but narrower than, argumentation-based reasoning and defeasible logic. Abstract argumentation studies how conflicting arguments attack one another and how acceptable argument sets can be defined \cite{dung1995acceptability}. Defeasible logic similarly represents rules whose conclusions may be overridden by stronger rules or priority relations \cite{nute1994defeasible}. The proposed implementation borrows this intuition through explicit conflicts and priorities, but it does not attempt to implement a full argumentation semantics. The design choice is pragmatic: the goal is to support reproducible policy-controller revision in simulation, not to contribute a new formal argumentation framework.

Finally, the approach differs from classical expert systems and rule-based controllers. Expert systems encode domain knowledge as rules, and rule-based controllers use conditional logic to select actions \cite{grosan2011rule}. A machine-coached controller adds an additional requirement: the rule base is not fixed after design. It is revised after diagnosed failures, and the revision is recorded as an explicit operation such as adding a rule, removing a rule, or changing a priority. Thus, the novelty is not the use of rules alone, but the closed loop:

\begin{quote}
	diagnostic failure $\rightarrow$ explanation of policy action $\rightarrow$ structured coaching advice $\rightarrow$ rule-base revision $\rightarrow$ held-out re-evaluation.
\end{quote}

For this reason, the term machine coaching is used to describe a constrained form of explanation-driven rule revision. The claim is not that the system performs unrestricted human coaching. The claim is that adaptive ABM policy controllers can be made more contestable when diagnostic evidence is connected to explicit, auditable revisions of the controller's policy logic \cite{garrone2025framework,garrone2026policy,arrieta2020explainable}.

\subsection{Machine coaching as controller-level contestability}

Within this paper, machine coaching is defined as a constrained revision protocol for an explainable policy controller. A coach may be a human expert, a stakeholder-facing analyst, or a scripted diagnostic module. In all cases, coaching is restricted to a small set of auditable operations on the controller's symbolic rule base \cite{grosan2011rule,nute1994defeasible}.

The coach does not provide unrestricted natural-language advice. Instead, coaching is represented as one of three revision operations:
\begin{enumerate}[label=(\alph*)]
	\item add a rule;
	\item remove a rule;
	\item change a rule priority.
\end{enumerate}

This restriction is intentional. It prevents the coaching layer from becoming an informal natural-language interface and preserves reproducibility. It also distinguishes the approach from ad hoc controller tuning. A revision counts as machine coaching in the present framework only if four conditions hold:

\begin{enumerate}[label=(\roman*)]
	\item a diagnostic failure is identified by a predefined criterion;
	\item the controller produces an explanation identifying the facts, rules, conflicts, and priorities involved in the policy decision;
	\item the coaching operation modifies the explicit rule base rather than directly editing the numerical outcome;
	\item the revised controller is re-evaluated on held-out seeds or cloned initial states.
\end{enumerate}

Thus, coaching is not defined by the mere presence of rules. It is defined by the link between explanation, contestation, revision, and re-evaluation \cite{arrieta2020explainable}. A rule-based controller without revision is simply a symbolic controller \cite{grosan2011rule}. A manually retuned controller without explanation logs is controller engineering. A machine-coached controller, as understood here, is a symbolic controller whose reasoning can be challenged and revised through explicit, logged operations after diagnosed failure modes \cite{nute1994defeasible,dung1995acceptability}.

	% ------------------------------------------------------------
	\section{Stylized Emissions-Regulation Experimental Environment}
	% ------------------------------------------------------------
	
	\subsection{Model overview}
	
	The empirical component of the article uses a stylized emissions-regulation ABM. It is not intended to forecast a real industrial sector. Its purpose is to provide a controlled environment in which policy-controller logic can be diagnosed and revised.
	
	There are $N$ firms. At time $t$, firm $i$ produces emissions $e_{i,t}$. Aggregate emissions are
	\begin{equation}
		E_t = \sum_{i=1}^{N} e_{i,t}.
	\end{equation}
	
	A regulator imposes an emissions cap $C$:
	\begin{equation}
		E_t \leq C.
	\end{equation}
	
	Regulatory pressure is represented by a scalar policy signal $P_t$, interpretable as a carbon tax, inspection pressure, compliance signal, or equivalent regulatory intensity.
	
	A simple firm-level emissions function is:
	\begin{equation}
		e_{i,t} =
		\max \left\{0,\ b_i - \eta_{i,t}P_t + \epsilon_{i,t} \right\},
	\end{equation}
	where $b_i$ is baseline emissions, $\eta_{i,t}$ is policy responsiveness, and $\epsilon_{i,t}$ is stochastic noise.
	
	Agent adaptation can be represented through an update of $\eta_{i,t}$:
	\begin{equation}
		\eta_{i,t+1}
		=
		\eta_{i,t}
		+
		\alpha_i g(P_t, E_t, C)
		+
		\nu_{i,t},
	\end{equation}
	where $\alpha_i$ is an adaptation rate, $g(\cdot)$ is a bounded response function, and $\nu_{i,t}$ is an idiosyncratic disturbance.
	
	\subsection{Numeric adaptive controllers}
	
	The baseline adaptive controllers are numeric. They update $P_t$ directly from observed emissions.
	
	A setpoint controller may be written as
	\begin{equation}
		P_{t+1} =
		\left[
		P_t + \kappa(E_t - C)
		\right]_{P_{\min}}^{P_{\max}},
	\end{equation}
	where $\kappa$ is a gain parameter and $[\cdot]_{P_{\min}}^{P_{\max}}$ denotes clipping.
	
	A safety-margin controller targets a stricter internal cap $C - m$:
	\begin{equation}
		P_{t+1} =
		\left[
		P_t + \kappa(E_t - (C-m))
		\right]_{P_{\min}}^{P_{\max}}.
	\end{equation}
	
	A one-sided controller increases pressure only when emissions exceed the cap:
	\begin{equation}
		P_{t+1} =
		\begin{cases}
			\left[P_t + \kappa(E_t-C)\right]_{P_{\min}}^{P_{\max}}, & E_t > C,\\
			P_t, & E_t \leq C.
		\end{cases}
	\end{equation}
	
	The one-sided controller is useful because it can reduce violations, but it may become over-conservative if it never relaxes pressure after emissions remain safely below the cap.
	
	% ------------------------------------------------------------
	\section{Symbolic Policy Controller}
	% ------------------------------------------------------------
	
	\subsection{Predicate abstraction}
	
	The symbolic controller does not consume all raw simulation variables. Instead, a predicate encoder maps numerical state variables into interpretable facts. Let
	\begin{equation}
		\mathcal{F}_t = A(s_t)
	\end{equation}
	be the set of facts extracted from state $s_t$ by abstraction function $A$.
	
	Examples include:
	\begin{itemize}
		\item \texttt{emissions\_above\_cap};
		\item \texttt{emissions\_near\_cap};
		\item \texttt{emissions\_safely\_below\_cap};
		\item \texttt{violations\_persistent};
		\item \texttt{policy\_volatility\_high};
		\item \texttt{conservatism\_gap\_high};
		\item \texttt{overshoot\_high}.
	\end{itemize}
	
	For example:
	\begin{equation}
		\texttt{emissions\_above\_cap}
		\in \mathcal{F}_t
		\quad \Longleftrightarrow \quad
		E_t > C.
	\end{equation}
	
	Similarly,
	\begin{equation}
		\texttt{emissions\_safely\_below\_cap}
		\in \mathcal{F}_t
		\quad \Longleftrightarrow \quad
		E_t < C - \delta_C,
	\end{equation}
	where $\delta_C$ is a safety threshold.
	
	\subsection{Rules, conflicts, and priorities}
	
	A policy rule is defined as follows.
	
	\begin{definition}[Policy rule]
		A policy rule is a tuple
		\begin{equation}
			r = (\mathrm{name}, B, h, \pi),
		\end{equation}
		where $\mathrm{name}$ is a rule identifier, $B$ is a finite set of body predicates, $h$ is a head or conclusion, and $\pi \in \mathbb{Z}$ is a priority score.
	\end{definition}
	
	A rule $r$ is applicable at time $t$ if
	\begin{equation}
		B_r \subseteq \mathcal{F}_t.
	\end{equation}
	
	Rules may support conflicting conclusions. For example:
	\begin{equation}
		\texttt{increase\_pressure}
		\;\bot\;
		\texttt{relax\_pressure}.
	\end{equation}
	
	Let $\mathcal{C}$ denote the conflict relation among rule heads. A selected rule set must be conflict-consistent according to $\mathcal{C}$ and the priority relation.
	
			A generic symbolic rule base may include rules of the following form:
			
	\begin{lstlisting}
		r1: IF emissions_above_cap
		THEN increase_pressure
		PRIORITY 1
		
		r2: IF emissions_near_cap AND violations_persistent
		THEN increase_pressure
		PRIORITY 2
		
		r3: IF policy_volatility_high
		THEN smooth_policy
		PRIORITY 3
		
		r4: IF emissions_safely_below_cap AND conservatism_gap_high
		THEN relax_pressure
		PRIORITY 2
		
		r5: IF emissions_above_cap AND overshoot_high
		THEN increase_pressure_strongly
		PRIORITY 4
	\end{lstlisting}
	
The list above should be read as the controller's admissible rule vocabulary rather than as
the exact initial rule base used in the controlled experiment. In the experiment reported below,
the initial symbolic controller deliberately excludes the relaxation rule. This omission creates
a transparent over-conservatism defect: when emissions are safely below the cap and the
conservatism gap is high, the controller has no exception allowing it to relax pressure. The
machine-coaching layer is then tested on whether it can diagnose and repair precisely this
missing policy-reasoning exception.

	\subsection{Decision procedure}
	
	At each time step, the controller performs four operations:
	\begin{enumerate}[label=(\roman*)]
		\item encode the state into facts;
		\item identify applicable rules;
		\item select a conflict-consistent subset using priorities;
		\item translate selected conclusions into a policy action.
	\end{enumerate}
	
	Let $\mathcal{R}$ be the full rule base and $\mathcal{R}_t^{app}$ the applicable rule set at time $t$:
	\begin{equation}
		\mathcal{R}_t^{app}
		=
		\{r \in \mathcal{R}: B_r \subseteq \mathcal{F}_t\}.
	\end{equation}
	
	The selected rule set $\mathcal{R}_t^{sel}$ is obtained by priority-ordered conflict resolution:
	\begin{equation}
		\mathcal{R}_t^{sel}
		=
		\operatorname{Resolve}(\mathcal{R}_t^{app}, \mathcal{C}, \pi).
	\end{equation}
	
	The policy action is then
	\begin{equation}
		a_t = H(\mathcal{R}_t^{sel}),
	\end{equation}
	where $H$ maps selected rule heads to numerical policy updates.
	
	For example:
	\begin{equation}
		P_{t+1} =
		\begin{cases}
			P_t + \Delta^+, & a_t = \texttt{increase\_pressure},\\
			P_t + \Delta^{++}, & a_t = \texttt{increase\_pressure\_strongly},\\
			P_t - \Delta^-, & a_t = \texttt{relax\_pressure},\\
			(1-\lambda)P_t + \lambda P_{t-1}, & a_t = \texttt{smooth\_policy}.
		\end{cases}
	\end{equation}
	
	\subsection{Priority-based rule resolution}
	
	The operator $\operatorname{Resolve}(\cdot)$ is implemented as a deterministic priority-based rule-resolution procedure. This choice is deliberately simple: the purpose is not to implement a full argumentation semantics, but to obtain a reproducible and auditable controller decision from a finite rule base.
	
	Each rule $r$ has a priority $\pi(r)$ and a body $B_r$. Priorities are treated as integer scores. Higher values indicate stronger rules. Priorities are static during a simulation episode and can change only through an explicit coaching operation. To avoid nondeterminism, ties are resolved by a fixed ordering key:

	\begin{equation}
		\operatorname{key}(r)
		=
		\left(
		\pi(r),\ |B_r|,\ -\operatorname{index}(r)
		\right)
	\end{equation}
	
	where $|B_r|$ is rule specificity and $\operatorname{index}(r)$ is the insertion order of the rule in the rule base. Rules are sorted in descending order of $\pi(r)$, then descending order of specificity, then ascending insertion order. Thus, if two conflicting rules have equal priority, the more specific rule is selected; if they are equally specific, the earlier rule is selected and the later rule is blocked. This convention makes the priority relation operationally total, even if the explicit priority scores alone are not.
	
	Let $\mathcal{C} \subseteq H \times H$ denote a symmetric conflict relation over rule heads. Two heads $h$ and $h'$ are incompatible if $(h,h') \in \mathcal{C}$ or $(h',h) \in \mathcal{C}$. A selected rule set is conflict-consistent if no pair of selected rule heads conflicts:
	
	\begin{equation}
		\forall r,r' \in \mathcal{R}^{sel}_t,\ r \neq r':
		(h_r,h_{r'}) \notin \mathcal{C}
		\ \wedge\
		(h_{r'},h_r) \notin \mathcal{C}.
	\end{equation}

	Multiple rules may be selected at the same decision step when their heads do not conflict. For example, \texttt{increase\_pressure} and \texttt{smooth\_policy} may be selected together if the conflict map does not declare them incompatible. In that case, the first head determines the direction of the policy update, while the second acts as a modifier of the numerical update. By contrast, \texttt{increase\_pressure} and \texttt{relax\_pressure} are mutually exclusive and cannot both be selected.
	
	Blocked rules are not discarded silently. Each blocked rule is recorded together with the selected rule that blocked it and the reason for blocking. This record becomes part of the explanation object returned by the controller.
	
	\begin{algorithm}[h!]
		\caption{Priority-Based Rule Resolution}
		\label{alg:rule-resolution}
		\begin{algorithmic}[1]
			\Require Facts $\mathcal{F}_t$, rule base $\mathcal{R}$, conflict relation $\mathcal{C}$
			\Ensure Selected rules $\mathcal{R}^{sel}_t$, blocked rules $\mathcal{R}^{blk}_t$, explanation object $\mathcal{E}_t$
			
			\State $\mathcal{R}^{app}_t \gets \{r \in \mathcal{R}: B_r \subseteq \mathcal{F}_t\}$
			\State Sort $\mathcal{R}^{app}_t$ by descending $\pi(r)$, descending $|B_r|$, ascending insertion order
			\State $\mathcal{R}^{sel}_t \gets \emptyset$
			\State $\mathcal{R}^{blk}_t \gets \emptyset$
			
			\ForAll{$r \in \mathcal{R}^{app}_t$}
			\State $\operatorname{conflict} \gets \textbf{false}$
			\State $\operatorname{blocker} \gets \emptyset$
			\ForAll{$r' \in \mathcal{R}^{sel}_t$}
			\If{$(h_r,h_{r'}) \in \mathcal{C}$ or $(h_{r'},h_r) \in \mathcal{C}$}
			\State $\operatorname{conflict} \gets \textbf{true}$
			\State $\operatorname{blocker} \gets r'$
			\State \textbf{break}
			\EndIf
			\EndFor
			
			\If{$\operatorname{conflict} = \textbf{false}$}
			\State $\mathcal{R}^{sel}_t \gets \mathcal{R}^{sel}_t \cup \{r\}$
			\Else
			\State $\mathcal{R}^{blk}_t \gets \mathcal{R}^{blk}_t \cup \{(r,\operatorname{blocker},\text{``conflicting head''})\}$
			\EndIf
			\EndFor
			
			\State $\mathcal{E}_t \gets \{\mathcal{F}_t,\mathcal{R}^{app}_t,\mathcal{R}^{sel}_t,\mathcal{R}^{blk}_t,\mathcal{C}\}$
			\State \Return $\mathcal{R}^{sel}_t,\mathcal{R}^{blk}_t,\mathcal{E}_t$
		\end{algorithmic}
	\end{algorithm}
	
	The policy action is then obtained by applying the action mapper $H$ to the selected rule set:
	\begin{equation}
		a_t = H(\mathcal{R}^{sel}_t).
	\end{equation}
	When several compatible heads are selected, $H$ first identifies the primary policy-direction head, such as \texttt{increase\_pressure}, \texttt{relax\_pressure}, or \texttt{hold\_pressure}, and then applies any compatible modifier heads, such as \texttt{smooth\_policy}. If no primary action is selected, the controller defaults to \texttt{hold\_pressure}. This default is also recorded in the explanation object.

	\subsection{Explanation object}
	
	Each policy decision produces an explanation object:
	\begin{lstlisting}
		{
			"facts": [...],
			"applicable_rules": [...],
			"selected_rules": [...],
			"blocked_rules": [...],
			"conflicts": [...],
			"action": "...",
			"policy_before": P_t,
			"policy_after": P_t+1
		}
	\end{lstlisting}
	
	This object is essential. Without it, the symbolic controller is merely another rule-based controller. With it, the controller becomes inspectable and contestable.
	
	% ------------------------------------------------------------
	\section{Machine-Coaching Revision Layer}
	% ------------------------------------------------------------
	
	\subsection{Diagnostic failure modes}
	
	The coaching layer is triggered by diagnostic failures. The first version of the protocol
	considers three primary failure modes: boundary violations, over-conservatism, and policy
	volatility. It also includes one persistence subcase, repeated overshoot, which is treated as a
	priority-adjustment case rather than as a separate controller family.

	\begin{table}[h!]
		\centering
		\caption{Diagnostic failure modes used for machine-coached revision.}
		\begin{tabularx}{\textwidth}{lllY}
			\toprule
			Failure mode & Diagnostic condition & Typical controller & Interpretation \\
			\midrule
			Boundary violation &
			$\mathrm{ViolationRate} > \tau_v$ &
			Setpoint &
			Controller tracks the cap but crosses it too often. \\
			
			Over-conservatism &
			$\mathrm{ConservatismGap} > \tau_c$ &
			One-sided &
			Controller avoids violations but remains too far below the cap. \\
			
			Volatility &
			$\mathrm{PolicyVolatility} > \tau_p$ &
			Adaptive controllers &
			Controller reacts too sharply to short-run deviations. \\
			\bottomrule
		\end{tabularx}
	\end{table}
	
	These failure modes are selected because they correspond to interpretable regulatory concerns. They are not treated as a complete welfare function.
	
	\subsection{Advice representation}
	
	A coaching advice object is defined as follows.
	
	\begin{definition}[Coaching advice]
		A coaching advice object is a tuple
		\begin{equation}
			c = (op, r, r^\star, \pi^\star, q),
		\end{equation}
		where $op$ is a revision operation, $r$ is a new rule if applicable, $r^\star$ is a target rule if applicable, $\pi^\star$ is a new priority if applicable, and $q$ is a textual or symbolic reason.
	\end{definition}
	
	The allowed operations are:
	\begin{itemize}
		\item \texttt{add\_rule};
		\item \texttt{remove\_rule};
		\item \texttt{change\_priority}.
	\end{itemize}
	
	Example advice for over-conservatism:
	\begin{lstlisting}
		operation: add_rule
		rule:
		IF emissions_safely_below_cap AND conservatism_gap_high
		THEN relax_pressure
		PRIORITY 5
		reason:
		Persistent safe emissions indicate excessive policy pressure.
	\end{lstlisting}
	
	Example advice for repeated violations:
	\begin{lstlisting}
		operation: add_rule
		rule:
		IF emissions_near_cap AND violation_trend_increasing
		THEN increase_pressure_preemptively
		PRIORITY 5
		reason:
		Waiting until emissions exceed the cap creates repeated violations.
	\end{lstlisting}

	\subsection{Predefined advice templates}
	
	To avoid treating the coach as an oracle, coaching advice is not generated freely after inspecting the full results. Instead, the first implementation uses a finite set of predefined advice templates. Each template maps a diagnosed controller failure to a fixed revision operation. The templates are specified before the held-out evaluation phase and are applied mechanically when their triggering conditions are satisfied.
	
	Let $D_k$ denote the diagnostic summary computed after a diagnosis run or batch. The coaching function is defined as
	\begin{equation}
		c_k = \Gamma(D_k,\mathcal{E}_k,\mathcal{R}_k),
	\end{equation}
	where $\Gamma$ is a predefined advice-template function, $\mathcal{E}_k$ is the set of explanation logs associated with the diagnosed failure, and $\mathcal{R}_k$ is the current rule base. The function $\Gamma$ is not allowed to invent arbitrary new advice during evaluation. It can only return one of the templates listed in Table~\ref{tab:advice-templates}, or return \texttt{no\_revision} if no template condition is satisfied.

	\begin{table}[htbp]
		\centering
		\caption{Predefined advice templates used by the coaching layer.}
		\label{tab:advice-templates}
		\scriptsize
		\setlength{\tabcolsep}{3.2pt}
		\renewcommand{\arraystretch}{1.18}
		
		\begin{threeparttable}
			\begin{tabular}{
					@{}
					>{\raggedright\arraybackslash}p{2.2cm}
					>{\raggedright\arraybackslash}p{2.3cm}
					>{\raggedright\arraybackslash}p{3.8cm}
					>{\raggedright\arraybackslash}p{4.3cm}
					@{}
				}
				\toprule
				Failure mode & Trigger condition & Revision operation & Intended effect \\
				\midrule
				
				Boundary violations
				&
				$V > \tau_v$ and $O > \tau_o$
				&
				Add rule:\par
				\texttt{IF emissions\_near\_cap AND}\par
				\texttt{\quad violation\_trend\_increasing}\par
				\texttt{THEN increase\_pressure}\par
				\texttt{\quad preemptively}\par
				with priority $\pi_p$
				&
				Reduce repeated cap crossing by acting before emissions exceed the cap. \\
				
				\addlinespace[0.35em]
				
				Over-conservatism
				&
				$CG > \tau_c$ and $V \leq \tau_v$
				&
				Add rule:\par
				\texttt{IF emissions\_safely\_below\_cap}\par
				\texttt{\quad AND conservatism\_gap\_high}\par
				\texttt{THEN relax\_pressure}\par
				with priority $\pi_r$
				&
				Reduce excessive regulatory pressure when compliance is already stable. \\
				
				\addlinespace[0.35em]
				
				Policy volatility
				&
				$\mathrm{PV} > \tau_p$
				&
				Add rule:\par
				\texttt{IF policy\_volatility\_high}\par
				\texttt{THEN smooth\_policy}\par
				with priority $\pi_s$
				&
				Reduce sharp policy oscillations caused by short-run emissions fluctuations. \\
				
				\addlinespace[0.35em]
				
				Repeated overshoot
				&
				$O > \tau_o$ and $\mathrm{MOL} > \tau_\ell$
				&
				Increase priority of:\par
				\texttt{increase\_pressure\_strongly}\par
				to $\pi_{++}$
				&
				Prioritize stronger correction when above-cap episodes persist. \\
				
				\addlinespace[0.35em]
				
				No diagnosed failure
				&
				All failure conditions false
				&
				\texttt{no\_revision}
				&
				Preserve the current controller. \\
				
				\bottomrule
			\end{tabular}
			
			\begin{tablenotes}[flushleft]
				\footnotesize
				\item \textit{Note.} $\mathrm{MOL}$ denotes mean overshoot episode length.
			\end{tablenotes}
		\end{threeparttable}
	\end{table}

	The thresholds $\tau_v$, $\tau_o$, $\tau_c$, $\tau_p$, and $\tau_\ell$ are fixed before the held-out evaluation. They may be chosen from a calibration set or specified as policy tolerances. The important methodological constraint is that they are not adjusted after observing the held-out results.
	
	This design separates machine coaching from manual controller redesign. A manually redesigned controller allows the analyst to inspect a failure and invent a new rule opportunistically. A template-based coached controller applies a predefined mapping from diagnostic evidence to rule-base revision. The coach therefore acts as a reproducible revision function rather than as an unconstrained source of new policy logic.

	\subsection{Revision operator}
	
	Let $\mathcal{R}_k$ denote the rule base before coaching event $k$. Let $c_k$ be the coaching advice. The revision operator is
	\begin{equation}
		\mathcal{R}_{k+1}
		=
		\operatorname{Revise}(\mathcal{R}_k, c_k).
	\end{equation}
	
	For the three allowed operations:
	\begin{equation}
		\operatorname{Revise}(\mathcal{R}, c) =
		\begin{cases}
			\mathcal{R} \cup \{r_c\}, & op(c)=\texttt{add\_rule},\\
			\mathcal{R} \setminus \{r^\star_c\}, & op(c)=\texttt{remove\_rule},\\
			(\mathcal{R}\setminus \{r^\star_c\}) \cup \{\tilde{r}^\star_c\}, & op(c)=\texttt{change\_priority}.
		\end{cases}
	\end{equation}
	
	The revised rule $\tilde{r}^\star_c$ is identical to $r^\star_c$ except that its priority is updated to $\pi^\star_c$.
	
	\subsection{Validity checks after coaching}
	
	A coaching operation is accepted only if the revised rule base passes a small set of syntactic and consistency checks. These checks are not intended to prove global optimality or logical completeness. They ensure that the controller remains executable and auditable after revision.
	
	Let $\mathcal{R}_{k+1}$ be the candidate rule base after applying coaching advice $c_k$. The revision is accepted only if:
	
	\begin{enumerate}[label=(\roman*)]
		\item every rule has a unique name;
		\item every body predicate belongs to the predefined predicate vocabulary;
		\item every rule head belongs to the predefined action or modifier vocabulary;
		\item the conflict relation $\mathcal{C}$ is symmetric;
		\item every action head that should be mutually exclusive with another action head is explicitly listed in $\mathcal{C}$;
		\item priorities are integer-valued;
		\item the rule-resolution algorithm returns a conflict-consistent selected set for all observed diagnostic states in the revision phase.
	\end{enumerate}
	
	Because the first implementation does not allow rules to generate new predicates recursively, rule chaining is bounded to one step:
	\[
	\mathcal{F}_t \rightarrow \mathcal{R}^{app}_t \rightarrow \mathcal{R}^{sel}_t \rightarrow a_t.
	\]
	Consequently, coaching cannot introduce inference cycles in the first version of the framework. Future extensions may allow intermediate conclusions or multi-step argumentation, but that would require additional cycle checks and a more explicit argumentation semantics.

	\subsection{Machine-coaching protocol}

		The full protocol is:
	
	\begin{enumerate}[
		label=\textbf{Step \arabic*.},
		widest=\textbf{Step 10.},
		leftmargin=*,
		labelsep=0.65em,
		align=left,
		itemsep=0.25em,
		topsep=0.35em
		]
		\item Run the ABM under an initial controller on diagnosis seeds.
		\item Compute scalar and symbolic diagnostics.
		\item Identify whether a predefined failure mode occurred.
		\item Retrieve the explanation logs associated with the failure.
		\item Apply the predefined advice-template function $\Gamma$.
		\item If $\Gamma$ returns \texttt{no\_revision}, preserve the current rule base.
		\item Otherwise, revise the rule base using the returned operation.
		\item Validate the revised rule base using the syntactic and consistency checks.
		\item Re-run the ABM from held-out seeds or cloned initial states.
		\item Compare recurrence of the diagnosed failure and guardrail metrics.
	\end{enumerate}

	The central restriction is that Step 5 is not discretionary during evaluation. The analyst does not freely generate new advice after observing the results. Advice is selected from a predefined template set. This makes the coaching layer reproducible and reduces the risk that apparent improvement is caused by post hoc controller tuning.

\begin{algorithm}[h!]
	\caption{Machine-coaching Advice}
	\label{alg:template-coach}
	\begin{algorithmic}[1]
		\Require Diagnostic summary $D_k$, explanation logs $\mathcal{E}_k$, rule base $\mathcal{R}_k$
		\Ensure Coaching advice $c_k$
		
		\If{$V(D_k) > \tau_v$ and $O(D_k) > \tau_o$}
		\State $c_k \gets$ \texttt{add\_rule(preemptive\_pressure)}
		\ElsIf{$CG(D_k) > \tau_c$ and $V(D_k) \leq \tau_v$}
		\State $c_k \gets$ \texttt{add\_rule(relaxation)}
		\ElsIf{$PV(D_k) > \tau_p$}
		\State $c_k \gets$ \texttt{add\_rule(smoothing)}
		\ElsIf{$O(D_k) > \tau_o$ and $\mathrm{MeanOvershootLength}(D_k) > \tau_\ell$}
		\State $c_k \gets$ \texttt{change\_priority(increase\_pressure\_strongly, $\pi_{++}$)}
		\Else
		\State $c_k \gets$ \texttt{no\_revision}
		\EndIf
		
		\State Attach supporting explanation logs from $\mathcal{E}_k$
		\State \Return $c_k$
	\end{algorithmic}
\end{algorithm}

	% ------------------------------------------------------------
	\section{Experimental Design}
	% ------------------------------------------------------------
	
	\subsection{Controller families}
	
	The experiment compares six controller families.
	
	\begin{table}[h!]
		\centering
		\caption{Controller families compared in the experimental protocol.}
		\begin{tabularx}{\textwidth}{llY}
			\toprule
			Code & Controller & Description \\
			\midrule
			F0 & Fixed policy & Constant policy signal. \\
			N1 & Numeric setpoint & Adjusts pressure according to cap deviation. \\
			N2 & Numeric safety-margin & Tracks an internal cap below the regulatory cap. \\
			N3 & Numeric one-sided & Increases pressure only when emissions exceed the cap. \\
			S1 & Symbolic rule-based & Uses explicit rules, conflicts, priorities, and explanations. \\
			S2 & Machine-coached symbolic & Revises symbolic rules after diagnostic failures. \\
			\bottomrule
		\end{tabularx}
	\end{table}
	
	\subsection{Regime comparison}
	
	The controller families are evaluated across four regimes:
	\begin{equation}
		\{CPCA, CPVA, VPCA, VPVA\}.
	\end{equation}
	
	The main focus is VPCA and VPVA because the coaching layer applies to adaptive policy controllers. CPCA and CPVA are retained as reference cases.
	
	\subsection{Training, coaching, and evaluation split}
	
	To avoid overfitting the controller to observed simulation failures, the experiment separates three phases.
	
	\begin{table}[h!]
		\centering
		\caption{Experimental phases.}
		\begin{tabularx}{\textwidth}{llY}
			\toprule
			Phase & Purpose & Data usage \\
			\midrule
			Diagnosis phase & Identify failure modes & Used to trigger coaching. \\
			Revision phase & Apply rule additions, removals, or priority changes & Uses only predefined advice templates. \\
			Evaluation phase & Test revised controller & Uses held-out seeds or cloned initial states. \\
			\bottomrule
		\end{tabularx}
	\end{table}
	
	This split is essential. Without it, improvements after coaching may reflect overfitting.
	
	\subsection{Primary evaluation criterion}
	
	The primary evaluation criterion is recurrence of the diagnosed failure mode:
	\begin{equation}
		\Delta R_f =
		R_f^{before} - R_f^{after},
	\end{equation}
	where $R_f$ is the recurrence rate of failure mode $f$ in held-out runs.
	
	This avoids claiming success from a single improved metric while ignoring deterioration elsewhere. The coached controller is considered successful for failure mode $f$ if $\Delta R_f > 0$ and the guardrail constraints defined below are satisfied. This prevents success from being defined after the fact by selecting whichever metric improves most.
	
	Guardrail metrics include:
	\begin{itemize}
		\item violation rate;
		\item overshoot;
		\item policy volatility;
		\item conservatism gap;
		\item mean tracking error.
	\end{itemize}

	\subsection{Predefined thresholds and guardrail tolerances}
	
	The evaluation protocol uses thresholds for two distinct purposes. First, diagnostic thresholds define when a controller failure is considered to have occurred and therefore when coaching is triggered. Second, guardrail tolerances define how much deterioration in secondary metrics is acceptable after coaching. Both sets of values are fixed before the held-out evaluation phase.
	
	Let $V$, $O$, $PV$, and $CG$ denote violation rate, mean overshoot, policy volatility, and conservatism gap, respectively. The diagnostic thresholds are:
	
	\begin{table}[h!]
		\centering
		\caption{Predefined diagnostic thresholds for coaching triggers.}
		\label{tab:diagnostic-thresholds}
		\begin{tabularx}{\textwidth}{lllY}
			\toprule
			Symbol & Metric & Suggested value & Interpretation \\
			\midrule
			$\tau_v$ & Violation rate & $0.05$ & Coaching is triggered if more than 5\% of time steps violate the cap. \\
			$\tau_o$ & Mean overshoot & $0.02C$ & Coaching is triggered if average overshoot exceeds 2\% of the emissions cap. \\
			$\tau_p$ & Policy volatility & $0.10(P_{\max}-P_{\min})$ & Coaching is triggered if mean absolute policy change exceeds 10\% of the feasible policy range. \\
			$\tau_c$ & Conservatism gap & $0.05C$ & Coaching is triggered if emissions remain, on average, more than 5\% of the cap below the safe operating region. \\
			$\tau_\ell$ & Mean overshoot episode length & $3$ time steps & Coaching is triggered if above-cap episodes persist for more than three consecutive steps on average. \\
			\bottomrule
		\end{tabularx}
	\end{table}
	
	These values are not presented as universal regulatory standards. They are operational thresholds for the stylized simulation environment. In applications, they should be replaced by domain-specific policy tolerances or selected through a calibration phase that is separated from held-out evaluation.
	
	The success of a coaching operation is evaluated by the reduction in recurrence of the diagnosed failure mode:
	\begin{equation}
		\Delta R_f =
		R_f^{before} - R_f^{after},
	\end{equation}
	where $R_f^{before}$ and $R_f^{after}$ denote the held-out recurrence rate of failure mode $f$ before and after coaching.
	
	A coaching operation is considered successful only if:
	\begin{equation}
		\Delta R_f > 0
	\end{equation}
	and no guardrail metric deteriorates beyond its predefined tolerance. Let $M_j^{before}$ and $M_j^{after}$ denote the value of guardrail metric $j$ before and after coaching. For metrics where lower values are preferred, deterioration is defined as:
	\begin{equation}
		\Delta M_j =
		M_j^{after} - M_j^{before}.
	\end{equation}
	The guardrail condition is:
	\begin{equation}
		\Delta M_j \leq \gamma_j
		\quad
		\forall j \in \mathcal{G},
	\end{equation}
	where $\mathcal{G}$ is the set of guardrail metrics and $\gamma_j$ is the maximum tolerated deterioration.
	
	\begin{table}[h!]
		\centering
		\caption{Guardrail tolerances for evaluating coached controllers.}
		\label{tab:guardrail-tolerances}
		\begin{tabularx}{\textwidth}{lllY}
			\toprule
			Symbol & Guardrail metric & Tolerance & Interpretation \\
			\midrule
			$\gamma_v$ & Violation rate & $0.01$ & Violation rate may not increase by more than one percentage point. \\
			$\gamma_o$ & Mean overshoot & $0.005C$ & Mean overshoot may not increase by more than 0.5\% of the cap. \\
			$\gamma_p$ & Policy volatility & $0.02(P_{\max}-P_{\min})$ & Policy volatility may not increase by more than 2\% of the feasible policy range. \\
			$\gamma_c$ & Conservatism gap & $0.01C$ & Conservatism gap may not increase by more than 1\% of the cap, unless conservatism is the diagnosed failure being corrected. \\
			$\gamma_{TE}$ & Tracking error & $0.01C$ & Mean tracking error may not increase by more than 1\% of the cap. \\
			\bottomrule
		\end{tabularx}
	\end{table}
	
	When a guardrail metric is itself the target of coaching, improvement in that metric is evaluated as the primary failure-reduction objective rather than as a guardrail. For example, if the diagnosed failure is over-conservatism, the primary objective is to reduce $CG$; in that case, $V$, $O$, $PV$, and $TE$ serve as guardrails.
	
	This explicit thresholding prevents flexible ex post success definitions. A coached controller is not judged successful merely because one favourable metric improves. It must reduce the predefined failure recurrence while remaining within the predefined guardrail tolerances.

\subsection{Simulation configuration}

Table~\ref{tab:simulation-configuration} reports the configuration used for the controlled simulation experiment. The table is included to make the numerical results reproducible from the manuscript. The experiment uses separate diagnosis and held-out evaluation seeds. Coaching advice is selected only from the diagnosis phase, while performance is evaluated on held-out seeds.

\begin{table}[htbp]
	\centering
	\caption{Simulation configuration for the controlled coaching experiment.}
	\label{tab:simulation-configuration}
	\scriptsize
	\setlength{\tabcolsep}{5pt}
	\renewcommand{\arraystretch}{1.12}
	
	\begin{tabular}{
			@{}
			>{\raggedright\arraybackslash}p{4.5cm}
			>{\centering\arraybackslash}p{1.6cm}
			l
			@{}
		}
		\toprule
		Parameter & Symbol & Value \\
		\midrule
		
		\multicolumn{3}{@{}l}{\textit{Simulation setup}} \\
		\addlinespace[0.15em]
		Number of firms & $N$ & 100 \\
		Simulation horizon & $T$ & 100 \\
		Burn-in & -- & 20 \\
		Emissions cap & $C$ & 75.0 \\
		Minimum policy signal & $P_{\min}$ & 0.0 \\
		Maximum policy signal & $P_{\max}$ & 25.0 \\
		Initial policy signal & $P_0$ & 18.0 \\
		
		\addlinespace[0.4em]
		\midrule
		
		\multicolumn{3}{@{}l}{\textit{Random-seed design}} \\
		\addlinespace[0.15em]
		Diagnosis seeds & -- & 1000--1039 \\
		Number of diagnosis seeds & -- & 40 \\
		Evaluation seeds & -- & 2000--2039 \\
		Number of evaluation seeds & -- & 40 \\
		
		\addlinespace[0.4em]
		\midrule
		
		\multicolumn{3}{@{}l}{\textit{Diagnostic thresholds}} \\
		\addlinespace[0.15em]
		Violation threshold & $\tau_v$ & 0.05 \\
		Overshoot threshold & $\tau_o$ & 1.50 \\
		Policy-volatility threshold & $\tau_p$ & 2.50 \\
		Conservatism threshold & $\tau_c$ & 3.75 \\
		Overshoot-length threshold & $\tau_\ell$ & 3.0 \\
		
		\addlinespace[0.4em]
		\midrule
		
		\multicolumn{3}{@{}l}{\textit{Acceptance guardrails}} \\
		\addlinespace[0.15em]
		Violation guardrail & $\gamma_v$ & 0.01 \\
		Overshoot guardrail & $\gamma_o$ & 0.375 \\
		Policy-volatility guardrail & $\gamma_p$ & 0.50 \\
		Conservatism guardrail & $\gamma_c$ & 0.75 \\
		Tracking-error guardrail & $\gamma_{TE}$ & 0.75 \\
		
		\bottomrule
	\end{tabular}
\end{table}

	% ------------------------------------------------------------
	\section{Metrics}
	% ------------------------------------------------------------
	
	Let $R$ denote the number of replications and $T$ the number of time steps.
	
	\subsection{Violation rate}
	
	\begin{equation}
		V =
		\frac{1}{RT}
		\sum_{r=1}^{R}
		\sum_{t=1}^{T}
		\mathbb{I}(E_{r,t} > C).
	\end{equation}
	
	\subsection{Mean overshoot}
	
	\begin{equation}
		O =
		\frac{1}{RT}
		\sum_{r=1}^{R}
		\sum_{t=1}^{T}
		\max(0, E_{r,t}-C).
	\end{equation}
	
	\subsection{Tracking error}
	
	\begin{equation}
		TE =
		\frac{1}{RT}
		\sum_{r=1}^{R}
		\sum_{t=1}^{T}
		|E_{r,t}-C|.
	\end{equation}
	
	\subsection{Policy volatility}
	
	\begin{equation}
		PV =
		\frac{1}{R(T-1)}
		\sum_{r=1}^{R}
		\sum_{t=2}^{T}
		|P_{r,t}-P_{r,t-1}|.
	\end{equation}
	
	\subsection{Conservatism gap}
	
	\begin{equation}
		CG =
		\frac{1}{RT}
		\sum_{r=1}^{R}
		\sum_{t=1}^{T}
		\max(0, C - E_{r,t} - \delta_C).
	\end{equation}
	
	\subsection{Explanation and revision metrics}
	
	The machine-coaching layer also produces controller-level metrics:
	\begin{itemize}
		\item number of applicable rules per decision;
		\item number of selected rules per decision;
		\item number of blocked rules per decision;
		\item number of conflicts per run;
		\item number of coaching interventions;
		\item number of persistent rules after revision;
		\item recurrence rate of previously corrected failures.
	\end{itemize}
	
	These metrics do not replace policy-performance metrics. They measure auditability and revision dynamics.

	% ------------------------------------------------------------
	\section{Simulation Results}
	% ------------------------------------------------------------
	
	The purpose of the experiment is to test whether a predefined coaching template can convert a diagnosed symbolic-controller failure into an explicit rule-base revision and whether the revised controller reduces recurrence of that failure under held-out simulation seeds.
	
	\begin{figure}[h!]
		\centering
		\includegraphics[width=\textwidth]{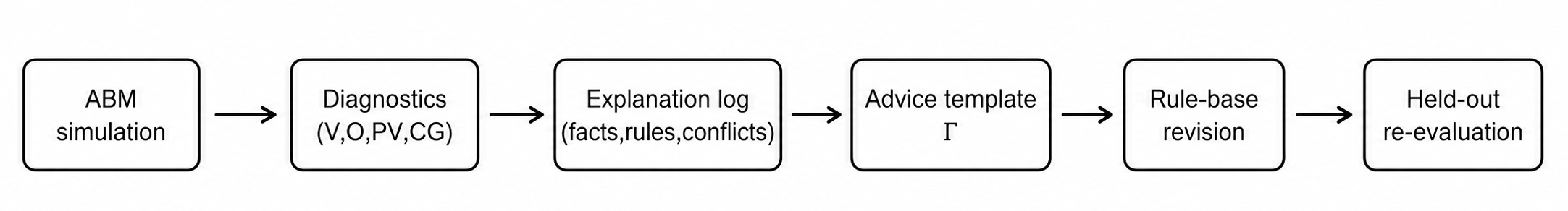}
		\caption{Machine-coaching controller revision loop. The workflow connects ABM simulation, scalar diagnostics, explanation logs, predefined advice templates, rule-base revision, and held-out re-evaluation.}
		\label{fig:coaching-architecture}
	\end{figure}
	
		The experiment should be interpreted as a controlled proof-of-concept rather than as broad empirical validation of machine coaching. The initial symbolic controller is intentionally specified without a relaxation exception. This creates a transparent controller-level defect: when emissions remain safely below the cap and the conservatism gap is high, the controller has no rule that can reduce excessive regulatory pressure. The purpose of the experiment is therefore not to show that machine coaching is generally superior to all adaptive-control alternatives, but to test whether the proposed architecture can expose a missing policy-reasoning exception, revise the rule base through a predefined template, and evaluate the revised controller on held-out seeds.
	
	The experiment focuses on the VPVA regime, where both the policy controller and the regulated agents are adaptive. The controlled failure mode is over-conservatism. In this setting, the uncoached symbolic controller maintains excessive regulatory pressure, keeping aggregate emissions safely below the cap but farther from the constraint than required. The predefined coaching function $\Gamma$ therefore selects the over-conservatism template and adds a relaxation rule to the symbolic controller.
	
		\begin{figure}[h!]
		\centering
		\includegraphics[width=\textwidth]{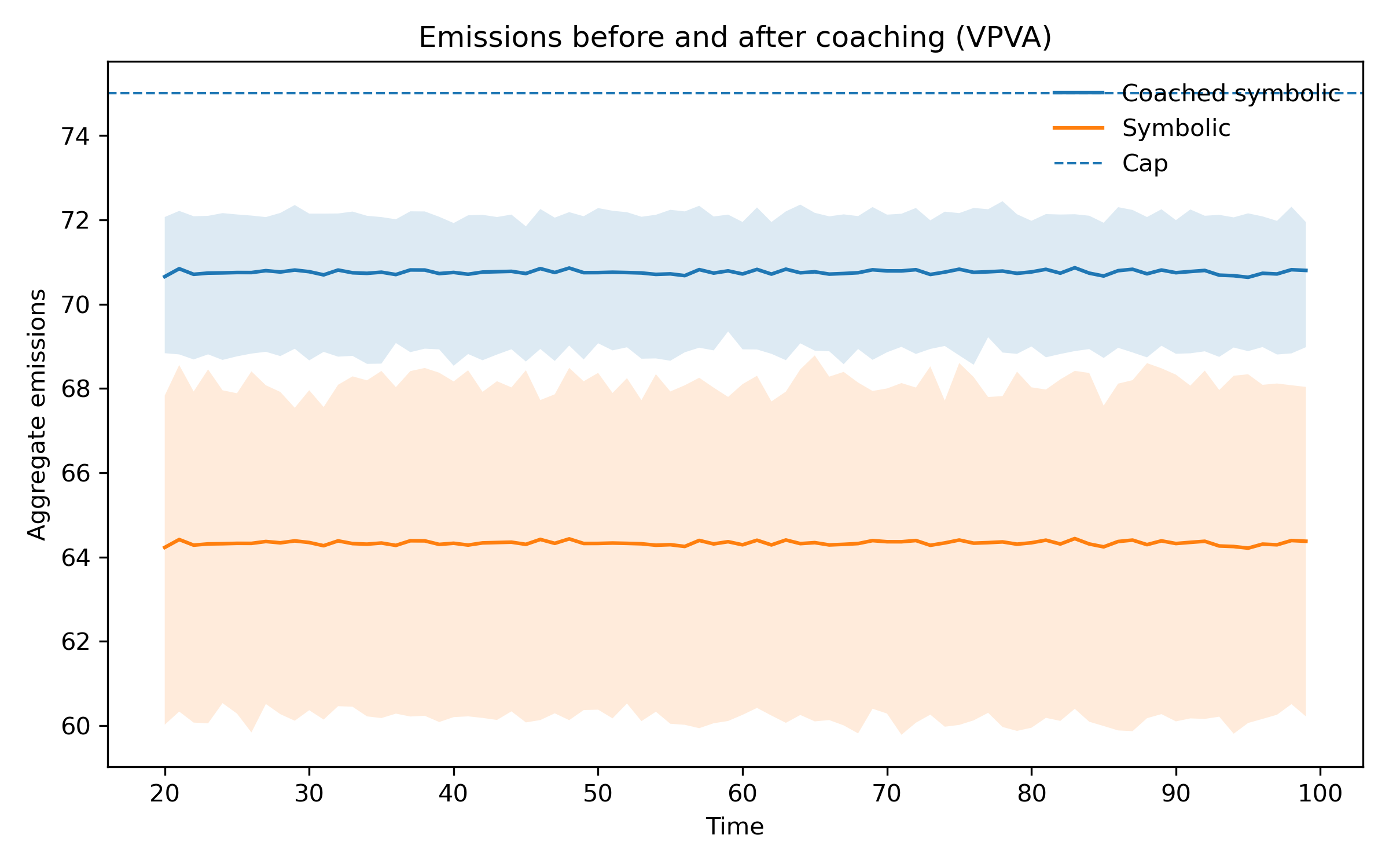}
		\caption{Aggregate emissions before and after coaching in the VPVA regime. The uncoached symbolic controller remains overly conservative, keeping emissions substantially below the cap. After template-constrained coaching, the added relaxation rule raises emissions closer to the cap while preserving compliance. Shaded areas indicate variability across held-out seeds.}
		\label{fig:emissions-before-after}
	\end{figure}
	
	\subsection{Diagnosis and coaching advice}
	
	Table~\ref{tab:diagnosis-summary} reports the diagnostic summary computed before coaching. The symbolic controller produces no boundary violations, no overshoot, and no policy volatility, but it exhibits a conservatism gap of $CG=6.574$ and a tracking error of $TE=10.324$. Since $CG > \tau_c$ and $V \leq \tau_v$, the predefined over-conservatism template is triggered.
	
	\begin{table}[h!]
		\centering
		\caption{Machine-coaching diagnosis summary before rule-base revision.}
		\label{tab:diagnosis-summary}
		\begin{tabular}{lrrrrrr}
			\toprule
			Controller & $V$ & $O$ & $PV$ & $CG$ & $TE$ & Mean overshoot length \\
			\midrule
			Symbolic & 0.000 & 0.000 & 0.000 & 6.574 & 10.324 & 0.000 \\
			\bottomrule
		\end{tabular}
	\end{table}
	
	The selected coaching advice is:
	
	\begin{quote}
		\texttt{add\_rule}: add \texttt{coach\_relax\_conservative}, with body
		\texttt{emissions\_safely\_below\_cap} and \texttt{conservatism\_gap\_high}, head
		\texttt{relax\_pressure}, and priority $5$.
	\end{quote}
	
	The reason attached to the advice is that persistent safe emissions indicate excessive policy pressure. This revision is not generated freely after inspecting the results. It is selected mechanically by the predefined advice-template function $\Gamma$.

\subsection{Uncertainty of the central before--after comparison}

Table~\ref{tab:paired-uncertainty} reports paired before--after differences over the 40 held-out evaluation seeds. The bootstrap confidence intervals are computed over the paired seed-level improvements. Positive values indicate improvement, since the reported difference is before minus after.

\begin{table}[htbp]
	\centering
	\caption{Paired held-out differences before and after coaching.}
	\label{tab:paired-uncertainty}
	\scriptsize
	\setlength{\tabcolsep}{3.5pt}
	\renewcommand{\arraystretch}{1.15}
	
	\begin{tabularx}{\textwidth}{
			>{\raggedright\arraybackslash}X
			S[table-format=2.3]
			S[table-format=1.3]
			S[table-format=1.3]
			S[table-format=1.3]
			S[table-format=1.3]
			>{\raggedleft\arraybackslash}p{2.15cm}
		}
		\toprule
		& \multicolumn{2}{c}{Before coaching}
		& \multicolumn{2}{c}{After coaching}
		& \multicolumn{2}{c}{Paired difference} \\
		\cmidrule(lr){2-3}
		\cmidrule(lr){4-5}
		\cmidrule(lr){6-7}
		Metric
		& {Mean}
		& {SD}
		& {Mean}
		& {SD}
		& {\makecell{Mean\\improvement}}
		& {\makecell{Bootstrap\\95\% CI}} \\
		\midrule
		Conservatism gap, $CG$
		& 6.952 & 2.998 & 0.770 & 1.014 & 6.182 & [5.073, 7.303] \\
		
		Tracking error, $TE$
		& 10.667 & 3.086 & 4.241 & 1.261 & 6.426 & [5.250, 7.608] \\
		
		Violation rate, $V$
		& 0.000 & 0.000 & 0.000 & 0.000 & 0.000 & [0.000, 0.000] \\
		
		Over-conservatism recurrence
		& 0.875 & 0.335 & 0.000 & 0.000 & 0.875 & [0.750, 0.975] \\
		\bottomrule
	\end{tabularx}
\end{table}

The paired comparison confirms that the improvement is not only visible in aggregate means. Across held-out seeds, coaching reduces both the conservatism gap and tracking error, while violation rate remains unchanged at zero. Over-conservatism recurrence decreases by 0.875, with a bootstrap 95\% confidence interval of [0.750, 0.975].

	\subsection{Controller summary table}
	
	Table~\ref{tab:vpva-controller-comparison} reports the VPVA controller-comparison results generated on held-out evaluation seeds. The main comparison is between the uncoached symbolic controller and the coached symbolic controller. Fixed-policy and numeric adaptive controllers are retained as interpretive baselines.
	
	\begin{table}[htbp]
		\centering
		\caption{VPVA controller comparison on held-out simulation seeds.}
		\label{tab:vpva-controller-comparison}
		\scriptsize
		\setlength{\tabcolsep}{4pt}
		\renewcommand{\arraystretch}{1.12}
		\begin{tabular}{
				@{}
				l
				S[table-format=1.3]
				S[table-format=1.3]
				S[table-format=2.3]
				S[table-format=1.3]
				S[table-format=1.3]
				S[table-format=1.3]
				r
				@{}
			}
			\toprule
			Controller 
			& {$V$} 
			& {$O$} 
			& {$\mathrm{TE}$} 
			& {$\mathrm{PV}$} 
			& {$CG$} 
			& {\makecell{Mean\\overshoot\\length}} 
			& {Runs} \\
			\midrule
			Fixed policy        & 0.000 & 0.000 & 10.667 & 0.000 & 6.952 & 0.000 & 40 \\
			One-sided           & 0.000 & 0.000 & 10.667 & 0.000 & 6.952 & 0.000 & 40 \\
			Safety-margin       & 0.000 & 0.000 & 3.772  & 0.019 & 0.137 & 0.000 & 40 \\
			Setpoint            & 0.431 & 0.101 & 0.260  & 0.020 & 0.000 & 1.718 & 40 \\
			Symbolic            & 0.000 & 0.000 & 10.667 & 0.000 & 6.952 & 0.000 & 40 \\
			Coached symbolic    & 0.000 & 0.000 & 4.241  & 0.000 & 0.770 & 0.000 & 40 \\
			\bottomrule
		\end{tabular}
	\end{table}

	The held-out VPVA comparison shows that the coached symbolic controller reduces the diagnosed over-conservatism failure. The uncoached symbolic controller has $CG=6.952$ and $TE=10.667$, while the coached symbolic controller has $CG=0.770$ and $TE=4.241$. Boundary violations, overshoot, and policy volatility remain equal to zero. Thus, the coached controller reduces over-conservatism without violating the predefined guardrails.
	
	The setpoint controller provides a useful contrast. It achieves low tracking error, but at the cost of frequent boundary violations, with a VPVA violation rate of $0.431$ and a mean overshoot of $0.101$. The safety-margin controller has the lowest conservatism gap among the non-symbolic adaptive controllers, but it is not explanation-driven and does not expose rule-level reasons for policy revision. The coached symbolic controller therefore occupies a different methodological position: its value lies not in numerical optimality, but in connecting diagnosis, explanation, revision, and held-out re-evaluation.
	
	\subsection{Failure recurrence table}
	
	Table~\ref{tab:failure-recurrence-results} reports failure recurrence before and after coaching. Over-conservatism recurrence falls from $0.875$ before coaching to $0.000$ after coaching. Boundary violations, policy-volatility failures, and repeated overshoot remain absent both before and after coaching.
	
	\begin{table}[h!]
		\centering
		\caption{Failure recurrence before and after coaching.}
		\label{tab:failure-recurrence-results}
		\begin{tabularx}{\textwidth}{lrrrY}
			\toprule
			Failure mode & Before coaching & After coaching & Change & Interpretation \\
			\midrule
			Boundary violations & 0.000 & 0.000 & 0.000 & Unchanged; no boundary-violation failure is introduced. \\
			Over-conservatism & 0.875 & 0.000 & 0.875 & Reduced; the diagnosed failure is eliminated in held-out evaluation. \\
			Policy volatility & 0.000 & 0.000 & 0.000 & Unchanged; the revision does not introduce volatility. \\
			Repeated overshoot & 0.000 & 0.000 & 0.000 & Unchanged; the revision does not introduce persistent above-cap episodes. \\
			\bottomrule
		\end{tabularx}
	\end{table}
	
	The result supports the restricted claim of the paper: the machine-coaching layer can convert an ex post diagnostic failure into a rule-level controller revision whose effect is measurable under held-out simulation seeds. It does not imply that machine coaching improves every controller metric or that the coached symbolic controller is globally optimal.

\subsection{Robustness checks}

Two local robustness checks are used to evaluate whether the result depends entirely on a single threshold or a single initial policy value. These checks are not intended to establish global robustness. They test whether the controlled proof-of-concept remains coherent under nearby conservatism thresholds and alternative initial policy pressures.

Table~\ref{tab:tau-c-sensitivity} reports sensitivity to the over-conservatism threshold $\tau_c$. The coaching template is triggered under all three tested values, $\tau_c \in \{0.03C,0.05C,0.07C\}$. The recurrence of over-conservatism decreases in all cases. Under the default threshold $\tau_c=0.05C$, recurrence falls from 0.875 to 0.000.

\begin{table}[h!]
	\centering
	\caption{Sensitivity to the over-conservatism threshold $\tau_c$.}
	\label{tab:tau-c-sensitivity}
	\scriptsize
	\begin{tabularx}{\textwidth}{llrrrrrr}
		\toprule
		$\tau_c$ & Advice & $\tau_c$ value & Before $CG$ & After $CG$ & Before recurrence & After recurrence & Change \\
		\midrule
		$0.03C$ & Over-conservatism & 2.25 & 6.952 & 0.770 & 0.950 & 0.125 & 0.825 \\
		$0.05C$ & Over-conservatism & 3.75 & 6.952 & 0.770 & 0.875 & 0.000 & 0.875 \\
		$0.07C$ & Over-conservatism & 5.25 & 6.952 & 0.770 & 0.750 & 0.000 & 0.750 \\
		\bottomrule
	\end{tabularx}
\end{table}

Table~\ref{tab:initial-policy-sensitivity} reports sensitivity to the initial policy signal $P_0$. Under the default batch-level diagnostic rule, coaching is not triggered for $P_0=14$ or $P_0=16$, because the diagnosis-phase conservatism gap does not exceed $\tau_c$. Coaching is triggered for $P_0=18$ and $P_0=20$, where the controller is substantially over-conservative. In both triggered cases, the coached controller reduces over-conservatism recurrence to zero.

\begin{table}[htbp]
	\centering
	\caption{Sensitivity to initial policy signal $P_0$.}
	\label{tab:initial-policy-sensitivity}
	\scriptsize
	\setlength{\tabcolsep}{3.5pt}
	\renewcommand{\arraystretch}{1.15}
	
	\begin{tabular}{
			@{}
			S[table-format=2.0]
			>{\raggedright\arraybackslash}p{2.8cm}
			S[table-format=2.3]
			S[table-format=2.3]
			S[table-format=2.3]
			S[table-format=1.3]
			S[table-format=1.3]
			S[table-format=1.3]
			S[table-format=2.3]
			S[table-format=1.3]
			@{}
		}
		\toprule
		& 
		& \multicolumn{3}{c}{Conservatism gap}
		& \multicolumn{3}{c}{Recurrence}
		& \multicolumn{2}{c}{Tracking error} \\
		\cmidrule(lr){3-5}
		\cmidrule(lr){6-8}
		\cmidrule(lr){9-10}
		
		{$P_0$}
		& {Advice}
		& {\makecell{Diagnosis\\$CG$}}
		& {\makecell{Before\\$CG$}}
		& {\makecell{After\\$CG$}}
		& {\makecell{Before}}
		& {\makecell{After}}
		& {\makecell{Change}}
		& {\makecell{Before\\$TE$}}
		& {\makecell{After\\$TE$}} \\
		\midrule
		
		14 & {No diagnosed failure}
		& 0.363 & 0.712 & 0.712
		& 0.050 & 0.050 & 0.000
		& 3.471 & 3.471 \\
		
		16 & {No diagnosed failure}
		& 2.700 & 3.242 & 3.242
		& 0.350 & 0.350 & 0.000
		& 6.844 & 6.844 \\
		
		18 & {Over-conservatism}
		& 6.574 & 6.952 & 0.770
		& 0.875 & 0.000 & 0.875
		& 10.667 & 4.241 \\
		
		20 & {Over-conservatism}
		& 10.556 & 10.915 & 0.241
		& 0.975 & 0.000 & 0.975
		& 14.665 & 3.567 \\
		
		\bottomrule
	\end{tabular}
\end{table}

These checks qualify the interpretation of the result. The coaching layer does not act continuously for every conservative configuration. It acts only when the predefined diagnostic condition is met. This is consistent with the design of the paper: the coach is a template-constrained revision function, not an unconstrained optimizer.

	\subsection{Explanation-log table}
	
	Table~\ref{tab:explanation-log-results} reports the explanation-log entry associated with the coached revision. Before coaching, the relevant facts are \texttt{emissions\_safely\_below\_cap} and \texttt{conservatism\_gap\_high}. The only applicable rule is the default hold-pressure rule. No rule is blocked. The absence of a relaxation rule explains why the symbolic controller remains conservative. The coaching template therefore adds a rule that relaxes policy pressure when emissions are safely below the cap and the conservatism gap is high.

\begin{table}[htbp]
	\centering
	\caption{Explanation log for the coached over-conservatism revision.}
	\label{tab:explanation-log-results}
	\scriptsize
	\setlength{\tabcolsep}{5pt}
	\renewcommand{\arraystretch}{1.15}
	
	\begin{tabular}{
			@{}
			>{\raggedright\arraybackslash}p{3.2cm}
			>{\raggedright\arraybackslash}p{8.8cm}
			@{}
		}
		\toprule
		Field & Example \\
		\midrule
		
		Detected failure 
		& Over-conservatism \\
		
		Facts 
		& \texttt{conservatism\_gap\_high}, 
		\texttt{emissions\_safely\_below\_cap} \\
		
		Applicable rule(s) 
		& \texttt{r0\_default\_hold} \\
		
		Blocked rule(s) 
		& none \\
		
		Action before revision 
		& \texttt{hold\_pressure} \\
		
		Coaching revision 
		& Add \texttt{coach\_relax\_conservative}:\par
		\texttt{IF emissions\_safely\_below\_cap}\par
		\texttt{AND conservatism\_gap\_high}\par
		\texttt{THEN relax\_pressure}\par
		with priority $5$. \\
		
		Expected effect 
		& Reduce excessive policy pressure while monitoring violations, 
		overshoot, policy volatility, and tracking error as guardrails. \\
		
		\bottomrule
	\end{tabular}
\end{table}

	This table illustrates the intended role of the explanation object. The controller does not merely report a numerical outcome. It identifies the facts, applicable rules, and missing exception that justify the revision. The rule-base update is therefore inspectable and auditable.
	
	% ------------------------------------------------------------
	\section{Discussion}
	% ------------------------------------------------------------
	
	The controlled simulation results show how the proposed machine-coaching layer changes the role of diagnostics in adaptive ABM. Diagnostics are not used only as ex post summaries. They are used as triggers for explicit controller revision. In the reported VPVA experiment, the symbolic controller is diagnosed as over-conservative: it produces no violations and no overshoot, but keeps emissions unnecessarily far below the cap. The explanation log identifies the reason at the controller level: the applicable rule is the default hold-pressure rule, and no relaxation rule is available.
	
	 The experiment shows that, in a controlled over-conservatism case, an explanation-driven revision layer can identify a missing controller exception and add it through a predefined advice template. This is sufficient for the methodological claim advanced here: machine coaching operationalizes controller-level contestability by making policy reasoning inspectable, revisable, and testable.
	 
	 	The robustness checks strengthen, but do not generalize, the proof-of-concept. The result is stable across nearby over-conservatism thresholds and across the initial policy values for which over-conservatism is actually diagnosed. At lower initial policy values, coaching is not triggered, and the symbolic controller remains unchanged. This behaviour is desirable in the present framework because the coach is designed as a bounded revision operator, not as a continuous optimizer.
	
	The predefined coaching template converts this diagnostic failure into a rule-level revision. The added rule, \texttt{coach\_relax\_conservative}, relaxes policy pressure when emissions are safely below the cap and the conservatism gap is high. Held-out evaluation then shows that over-conservatism recurrence falls from $0.875$ to $0.000$, while violations, overshoot, and policy volatility remain at zero. This provides a compact demonstration of controller-level contestability: the controller's reasoning is explained, challenged, revised, and re-evaluated.
	
	The coached symbolic controller is not shown to be globally optimal. In fact, the numeric setpoint controller achieves very low tracking error but does so with frequent cap violations. The safety-margin controller achieves a lower conservatism gap than the coached symbolic controller in this simulation run, but it does not provide the same rule-level explanation and revision trace. The contribution of the coached controller is therefore not numerical dominance across all metrics. Its contribution is that a diagnosed failure can be linked to an auditable policy-reasoning defect and corrected through a predefined revision operation.

	This distinction is important. The machine-coaching layer does not explain the full ABM. It does not identify all causal mechanisms behind emergent macro-patterns. It does not replace sensitivity analysis, structural causal modeling, computational-mechanics diagnostics, trajectory mining, or clustering. It explains and revises only the policy-controller logic. In this sense, the proposed layer is complementary to broader ABM explainability methods.
	
	The approach is also not equivalent to full human-in-the-loop machine coaching. In the present implementation, coaching is constrained to predefined advice templates. This restriction improves reproducibility but limits expressiveness. Future work may extend the protocol to expert elicitation, stakeholder argumentation, or natural-language interfaces. Such extensions would introduce additional validation problems, including how advice is elicited, how conflicting stakeholder claims are resolved, and how coached revisions are protected against strategic or biased intervention.

	% ------------------------------------------------------------
	\section{Limitations}
	% ------------------------------------------------------------
	
	The first limitation is that the emissions model is stylized. It is a controlled simulation environment, not a calibrated emissions-sector model. Numerical results should therefore be interpreted as evidence about controller behavior under controlled assumptions, not as environmental-policy forecasts.
	
	The second limitation is that the empirical demonstration isolates a single failure mode. The reported experiment focuses on over-conservatism in the VPVA regime. The coached controller stabilizes at a lower policy level after adding the relaxation rule, thereby reducing the conservatism gap. This is a proof-of-concept for explanation-driven rule revision, not evidence that the same coaching layer will improve all adaptive policy failures. A related limitation is that the empirical result is partly constructed by design. The initial symbolic controller lacks a relaxation rule, the diagnostic procedure detects the resulting over-conservatism, and the predefined coaching template adds the missing relaxation rule. This design is appropriate for a proof-of-concept because it creates a transparent and auditable controller defect. However, it does not establish that the same coaching architecture will discover non-obvious revisions, improve controllers under ambiguous failures, or outperform well-tuned numerical controllers.
	
	The third limitation concerns the symbolic abstraction. Indeed, predicate thresholds such as \texttt{emissions\_near\_cap}, \texttt{emissions\_safely\_below\_cap}, and \texttt{conservatism\_gap\_high} are design choices. They must be documented and subjected to sensitivity analysis. Different threshold values may alter which advice template is triggered.
	
	The fourth limitation concerns coaching. Although the present implementation avoids free-form advice by using predefined templates, the templates themselves are design choices. They should be fixed before held-out evaluation and, in applied settings, justified through domain expertise, policy tolerance, or a separate calibration phase.
	
	The fifth limitation is that rule-based explanations may be incomplete. A controller may explain why it selected a policy action, but that does not explain all downstream effects of the action in the ABM. Controller-level explainability and system-level explainability remain distinct.
	
	The sixth limitation is that better explainability may not imply better scalar performance. A coached symbolic controller may be more auditable while being less efficient under a narrow objective. This is not necessarily a failure if the purpose is contestable policy design rather than pure optimization. However, it limits the claim that can be made from the present experiment.

Last, the robustness checks are local. They vary the over-conservatism threshold and initial policy signal, but they do not exhaust the full parameter space of the emissions model, agent-adaptation dynamics, symbolic predicate definitions, or coaching-template design. They should therefore be interpreted as evidence that the controlled result is not tied to a single threshold value, not as evidence of global robustness.

	% ------------------------------------------------------------
	\section{Conclusion}
	% ------------------------------------------------------------
	
	This paper proposed and implemented a lightweight machine-coached policy-revision layer for adaptive agent-based regulation. The layer represents policy decisions as symbolic rules with conflicts and priorities, produces explanation logs for controller actions, and allows diagnostic failures to be translated into explicit rule revisions.
	
	The controlled simulation experiment demonstrates the mechanism on a specific VPVA over-conservatism case. The uncoached symbolic controller produces no cap violations but remains excessively conservative, with a held-out conservatism gap of $CG=6.952$ and tracking error of $TE=10.667$. The predefined coaching function selects the over-conservatism template and adds a relaxation rule. After coaching, the held-out conservatism gap falls to $CG=0.770$ and tracking error falls to $TE=4.241$, while violation rate, overshoot, and policy volatility remain at zero. Over-conservatism recurrence falls from $0.875$ to $0.000$.
	
	Machine coaching is not presented as a complete ABM explainability solution or as an optimal-control method. Instead, it is a controller-level contestability mechanism that complements existing diagnostic approaches. Its value lies in connecting ex post regime diagnostics to reproducible policy-controller revision.
	
	The proposed protocol is especially suitable as a derived extension of adaptive ABM research. In a broader explainability architecture, causal diagnostics, information-theoretic measures, trajectory mining, and clustering explain system dynamics; machine coaching explains and revises the controller that acts on those dynamics.

	% ------------------------------------------------------------
	\section*{Code and Data Availability}
	% ------------------------------------------------------------
	
The replication package includes the simulation extension, notebook integration code, configuration parameters, random seeds, diagnosis outputs, held-out evaluation outputs, controller-level metrics, explanation logs, rule-revision logs, and all tables and figures used in the manuscript. The repository will be archived with a persistent identifier before public release.

\section*{Disclosure on the Use of Generative AI}

Large language model tools were used to assist with language editing, LaTeX formatting, figure-caption refinement, and code-debugging support. All scientific content, simulation design, numerical results, interpretation, and final claims were reviewed and validated by the author, who remains fully responsible for the manuscript.

	% ------------------------------------------------------------
	% REFERENCES
	% ------------------------------------------------------------


\begin{thebibliography}{99}
		
	\bibitem{arthur1994}
	Arthur, W. B. (1994).
	Inductive reasoning and bounded rationality.
	\textit{American Economic Review}, 84(2), 406--411.
	\url{https://ideas.repec.org/a/aea/aecrev/v84y1994i2p406-11.html}
	
	\bibitem{busoniu2008}
	Busoniu, L., Babuska, R., \& De Schutter, B. (2008).
	A comprehensive survey of multiagent reinforcement learning.
	\textit{IEEE Transactions on Systems, Man, and Cybernetics, Part C}, 38(2), 156--172.
	\url{https://doi.org/10.1109/TSMCC.2007.913919}
	
	\bibitem{conte2014}
	Conte, R., \& Paolucci, M. (2014).
	On agent-based modeling and computational social science.
	\textit{Frontiers in Psychology}, 5, 668.
	\url{https://doi.org/10.3389/fpsyg.2014.00668}
	
	\bibitem{crutchfield1994}
	Crutchfield, J. P. (1994).
	The calculi of emergence: Computation, dynamics and induction.
	\textit{Physica D: Nonlinear Phenomena}, 75(1--3), 11--54.
	\url{https://doi.org/10.1016/0167-2789(94)90273-9}
	
	\bibitem{epstein1999}
	Epstein, J. M. (1999).
	Agent-based computational models and generative social science.
	\textit{Complexity}, 4(5), 41--60.
	\url{https://onlinelibrary.wiley.com/doi/10.1002/%28SICI%291099-0526%28199905/06%294%3A5%3C41%3A%3AAID-CPLX9%3E3.0.CO%3B2-F}
	
	\bibitem{epstein2012}
	Epstein, J. M. (2012).
	\textit{Generative Social Science: Studies in Agent-Based Computational Modeling}.
	Princeton University Press.
	\url{https://www.jstor.org/stable/j.ctt7rxj1}
	
	\bibitem{garrone2025framework}
	Garrone, R. (2025).
	An adaptive, data-integrated agent-based modeling framework for explainable and contestable policy design.
	\textit{arXiv preprint arXiv:2511.19726}.
	\url{https://arxiv.org/abs/2511.19726}
	
	\bibitem{garrone2026policy}
	Garrone, R. (2026).
	Structural distinguishability of static and adaptive policy regimes in agent-based regulation.
	Preprint.
	
	\bibitem{gilbert2008}
	Gilbert, N. (2008).
	\textit{Agent-Based Models}.
	SAGE Publications.
	\url{https://doi.org/10.4135/9781412983259}
	
	\bibitem{mccarthy1959}
	McCarthy, J. (1959).
	Programs with common sense.
	In \textit{Proceedings of the Teddington Conference on the Mechanization of Thought Processes}.
	\url{http://jmc.stanford.edu/articles/mcc59/mcc59.pdf}
	
	\bibitem{shalizi2001}
	Shalizi, C. R., \& Crutchfield, J. P. (2001).
	Computational mechanics: Pattern and prediction, structure and simplicity.
	\textit{Journal of Statistical Physics}, 104, 817--879.
	\url{https://doi.org/10.1023/A:1010388907793}
	
	\bibitem{tesfatsion2006}
	Tesfatsion, L. (2006).
	Agent-based computational economics: A constructive approach to economic theory.
	In L. Tesfatsion \& K. L. Judd (Eds.),
	\textit{Handbook of Computational Economics}, Vol. 2.
	Elsevier.
	\url{https://doi.org/10.1016/S1574-0021(05)02016-2}
	
	\bibitem{tesfatsionjudd2006}
	Tesfatsion, L., \& Judd, K. L. (Eds.). (2006).
	\textit{Handbook of Computational Economics, Volume 2: Agent-Based Computational Economics}.
	Elsevier.
	\url{https://shop.elsevier.com/books/handbook-of-computational-economics/tesfatsion/978-0-444-51253-6}
	
	\bibitem{zhang2021}
	Zhang, K., Yang, Z., \& Basar, T. (2021).
	Multi-agent reinforcement learning: A selective overview of theories and algorithms.
	In \textit{Handbook of Reinforcement Learning and Control}.
	Springer.
	\url{https://arxiv.org/abs/1911.10635}
	
	\bibitem{amershi2014power}
	Amershi, S., Cakmak, M., Knox, W. B., \& Kulesza, T. (2014).
	Power to the people: The role of humans in interactive machine learning.
	\textit{AI Magazine}, 35(4), 105--120.
	\url{https://doi.org/10.1609/aimag.v35i4.2513}
	
	\bibitem{arrieta2020explainable}
	Arrieta, A. B., D\'iaz-Rodr\'iguez, N., Del Ser, J., Bennetot, A., Tabik, S., Barbado, A.,
	Garc\'ia, S., Gil-L\'opez, S., Molina, D., Benjamins, R., Chatila, R., \& Herrera, F. (2020).
	Explainable artificial intelligence (XAI): Concepts, taxonomies, opportunities and challenges toward responsible AI.
	\textit{Information Fusion}, 58, 82--115.
	\url{https://doi.org/10.1016/j.inffus.2019.12.012}
	
	\bibitem{christiano2017deep}
	Christiano, P. F., Leike, J., Brown, T. B., Martic, M., Legg, S., \& Amodei, D. (2017).
	Deep reinforcement learning from human preferences.
	In \textit{Advances in Neural Information Processing Systems}.
	\url{https://arxiv.org/abs/1706.03741}
	
	\bibitem{dung1995acceptability}
	Dung, P. M. (1995).
	On the acceptability of arguments and its fundamental role in nonmonotonic reasoning, logic programming and n-person games.
	\textit{Artificial Intelligence}, 77(2), 321--357.
	\url{https://doi.org/10.1016/0004-3702(94)00041-X}
	
	\bibitem{grosan2011rule}
	Grosan, C., \& Abraham, A. (2011).
	Rule-based expert systems.
	In \textit{Intelligent Systems: A Modern Approach}.
	Springer.
	\url{https://link.springer.com/book/10.1007/978-3-642-21004-4}
	
	\bibitem{karimi2020algorithmic}
	Karimi, A.-H., Sch\"olkopf, B., \& Valera, I. (2021).
	Algorithmic recourse: From counterfactual explanations to interventions.
	In \textit{Proceedings of the ACM Conference on Fairness, Accountability, and Transparency}.
	\url{https://doi.org/10.1145/3442188.3445899}
	
	\bibitem{nute1994defeasible}
	Nute, D. (1994).
	Defeasible logic.
	In D. M. Gabbay, C. J. Hogger, \& J. A. Robinson (Eds.),
	\textit{Handbook of Logic in Artificial Intelligence and Logic Programming}, Vol. 3.
	Oxford University Press.
	\url{https://dl.acm.org/doi/10.5555/186124.186131}
	
	\bibitem{wachter2017counterfactual}
	Wachter, S., Mittelstadt, B., \& Russell, C. (2017).
	Counterfactual explanations without opening the black box: Automated decisions and the GDPR.
	\textit{Harvard Journal of Law \& Technology}, 31(2), 841--887.
	\url{https://arxiv.org/abs/1711.00399}
	
	\bibitem{michael2019machinecoaching}
	Michael, L. (2019).
	Machine Coaching.
	In \textit{Proceedings of the IJCAI 2019 Workshop on Explainable Artificial Intelligence (XAI)}.
	Macao, China.
	\url{https://www.researchgate.net/publication/334989337_Machine_Coaching}
	
	\bibitem{markos2022proxycoaches}
	Markos, V., Thoma, M., \& Michael, L. (2022).
	Machine Coaching with Proxy Coaches.
	In \textit{Proceedings of the Workshop on Argumentation and Machine Learning (ArgML@COMMA)}.
	CEUR Workshop Proceedings, Vol. 3208.
	\url{https://ceur-ws.org/Vol-3208/paper4.pdf}
	
		
	\end{thebibliography}
\end{document}